\documentclass[range]{ar2e}

\def\be{\begin{equation}}
\def\ee{\end{equation}}

\def\lsim{\mathrel{\raise.3ex\hbox{$<$\kern-.75em\lower1ex\hbox{$\sim$}}}}
\def\gsim{\mathrel{\raise.3ex\hbox{$>$\kern-.75em\lower1ex\hbox{$\sim$}}}}
\def\ifmath#1{\relax\ifmmode #1\else $#1$\fi}
\def\half{\ifmath{{\textstyle{1 \over 2}}}}

\def\to{\rightarrow}
\def\epem{e^+e^-}

\def\ma{m_a}
\def\mh{m_h}
\def\hi{h_1}

\def\ai{a_1}
\def\a{a}
\def\ma{m_{\a}}
\def\mhi{m_{h_1}}
 
\def\mai{m_{a_1}}
\def\mueff{\mu_{\rm eff}}
\def\mtau{m_\tau}
\def\noi{\noindent}

\def\br{{\rm Br}}
\def\tanb{\tan\beta}

\def\mstopbar{\overline m_{\wtil t}}
\def\mhusq{m_{H_u}^2}
\def\mhdsq{m_{H_d}^2}
\def\mssq{m_S^2}

\def\anti{\overline}

\def\call{{\cal L}}

\def\wtil{\widetilde}  
\def\what{\widehat}

\def\hsm{h_{\rm SM}}
\def\mhsm{m_{\hsm}}
   
\def\hh{H}

\def\mhh{m_{\hh}}

\def\tanb{\tan\beta}

\def\mb{m_b}
\def\mz{M_Z}
\def\mw{M_W}
\def\mgut{M_{\rm GUT}}

\def\akap{A_\kappa}
\def\alam{A_\lambda}
\def\cta{\cos\theta_A}
\def\sta{\sin\theta_A}

\def\ie{{\it i.e.}}  
\def\eg{{\it e.g.}}

\def\etc{{\it etc.}}

\newcommand{\nc}{\newcommand}
\nc{\beq}{\begin{equation}}   \nc{\eeq}{\end{equation}}
\nc{\bea}{\begin{eqnarray}}   \nc{\eea}{\end{eqnarray}}
\nc{\baa}{\begin{array}}      \nc{\eaa}{\end{array}}
\nc{\bit}{\begin{itemize}}    \nc{\eit}{\end{itemize}}
\nc{\ben}{\begin{enumerate}}  \nc{\een}{\end{enumerate}}
\nc{\bce}{\begin{center}}     \nc{\ece}{\end{center}}
\def\beqa{\begin{eqnarray}}
\def\eeqa{\end{eqnarray}}  
\def\bed{\begin{description}}
\def\eed{\end{description}}

\def\vev#1{\langle #1 \rangle}

\def\susy{{\rm SUSY}}

\def\gev{~{\rm GeV}}

\def\tev{~{\rm TeV}}

\def\fbi{~{\rm fb}^{-1}}
\def\hsm{h_{\rm SM}}
\def\mhsm{m_{\hsm}}
\newcommand{\met}{\mbox{${\rm \not\! E}_{\rm T}$}}
\newcommand{\me}{\mbox{${\rm \not\! E}$}}

\begin{document}



\jname{Annual Review of Nuclear Science}
\jyear{2008}
\jvol{1?}
\ARinfo{1056-8700/97/0610-00?}

\title{Nonstandard Higgs Boson Decays}

\markboth{Nonstandard Higgs Boson Decays}{Nonstandard Higgs Boson Decays}

\author{Spencer Chang
\affiliation{Center for Cosmology and Particle Physics, Dept. of
  Physics, New York University, New York, NY 10003; email: chang@physics.nyu.edu} 
Radovan Derm\' \i\v sek
\affiliation{School of Natural Sciences, Institute for Advanced Study,
  Princeton, NJ 08540; email: dermisek@ias.edu} 
John F. Gunion \affiliation{Department of Physics, University of
  California at Davis, Davis, CA 95616; email: gunion@physics.ucdavis.edu}
Neal Weiner
\affiliation{Center for Cosmology and Particle Physics, Dept. of
  Physics, New York University, New York, NY 10003; email: neal.weiner.nyu@gmail.com}}


\begin{abstract}
  This review summarizes the motivations for and phenomenological
  consequences of nonstandard Higgs boson decays, with emphasis on
  final states containing a pair of non-Standard-Model particles that
  subsequently decay to Standard Model particles. Typically these
  non-Standard-Model particles are part of a ``hidden'' sector, for
  example a pair of neutral Higgs bosons or a pair of
  unstable neutralinos. We emphasize that such decays
  allow for a Higgs substantially below the Standard Model Higgs LEP limit of
  114 GeV. This in turn means that the ``fine-tuning'' problems of
  many Beyond the Standard Model (BSM) theories, in particular
  supersymmetric models, can be eliminated while achieving excellent
  consistency with precision electroweak data which favor a Higgs
  boson with mass below $100\gev$ and standard $WW$, $ZZ$, and top couplings.
  
\end{abstract}

\maketitle

\section{INTRODUCTION}

Understanding the origin of the electroweak symmetry breaking (EWSB)
responsible for giving mass to the $W$ and $Z$ gauge bosons of the Standard Model (SM) is the
next major step in constructing the ultimate theory of particles and
their interactions. The Large Hadron Collider (LHC) is designed
specifically to explore the mechanism behind EWSB.  In particular, its
$14\tev$ center of mass energy and greater than $100 \fbi$ integrated luminosity 
is such that $WW\rightarrow WW$ scattering can be studied at energies
up to about $1\tev$ where unitarity would be violated if no new
physics associated with EWSB exists. 

While many mechanisms for EWSB have been explored, the simplest is the
introduction of one or more elementary spin-0 Higgs fields that
acquire vacuum expectation values (vevs) and which couple to $WW$ and
$ZZ$.  Then, the $W$ and $Z$ gauge bosons acquire a contribution to
their mass from each such Higgs field proportional to the strength of
the Higgs-$WW$ and Higgs-$ZZ$ coupling times the vev of the Higgs
field. The quantum fluctuation of the Higgs field relative to its vev
is a spin-0 particle called a Higgs boson. If the corresponding field
couples to $ZZ$ and $WW$ then so will the Higgs boson.

While the role of the Higgs vev is to give mass to the $W$ and $Z$, it
is the Feynman diagrams involving the Higgs bosons that prevent
unitarity violation in $WW$ scattering provided the Higgs bosons are
light enough --- roughly below $1\tev$.  If the Higgs bosons whose
corresponding fields have significant vevs are below about
$300\gev$, then $WW$ scattering will remain perturbative at all
energies.  Furthermore, precision measurements of the properties of $W$
and $Z$ gauge bosons are most consistent if the vev-weighted average
of the logarithms of the Higgs masses is somewhat below $\ln 100$ (GeV
units), i.e. 
\bea
\sum_i \frac{v_i^2}{v_{SM}^2} \ln m_{h_i} \leq \ln \left( 100 \gev \right),
\eea
where $\vev{\Phi_j}\equiv v_j$ and $\sum_j v_j^2=v_{SM}^2 \sim (175 \gev)^2$ is the square of
the vev of the Standard Model Higgs field.
Thus, a very attractive possibility is that 
Higgs bosons with significant $WW/ZZ$ couplings are rather light.

While a $100\gev$ Higgs mass is certainly acceptable within the
context of the renormalized SM, it requires an enormous cancellation
between the bare Higgs mass term in the vev-shifted Lagrangian and the
superficially quadratically divergent loop corrections to the mass,
especially that arising from the top quark loop.
This has led to the idea that there must be new physics below the TeV
scale that will regulate these quadratic divergences. For models with
elementary Higgs fields, supersymmetry (SUSY) is the earliest proposal for
such new physics and remains very attractive.   In supersymmetry, the loop
corrections containing superparticles come in with opposite sign with
respect to those with particles and the quadratic divergences are
canceled. If the mass of the stop and the masses of other
superparticles are below about $500\gev$, this cancellation will
result in a Higgs boson mass in the $100\gev$ range more or less
automatically.
A higher mass for the sparticles
would lead to a larger Higgs mass but, as discussed in later sections,
would lead to the need to ``fine-tune'' 
the soft SUSY breaking parameters at the GUT scale
in order that the correct value of the $Z$ boson mass is obtained. 
The equivalent problem in more general beyond the Standard Model
(BSM) theories would be the need to choose 
parameters at the new physics scale and/or coupling unification scale
with great precision in order to obtain the correct value of $\mz$.

The fine-tuning problem is closely related to the ``little hierarchy
problem'' which occurs in a wide variety of BSM theories, including
not only supersymmetric models but also Randall-Sundrum theories
\cite{Randall:1999ee} and Little Higgs theories
\cite{Schmaltz:2005ky,Perelstein:2005ka}. Sometimes referred to as the
``LEP Paradox'' \cite{Barbieri:2000gf}, there is a basic tension
existing at present in BSM physics. On the one hand, precision electroweak
fits show no need for physics beyond the Standard Model,~\footnote{A
  possible exception is the 2.2 $\sigma$ discrepancy in $g-2$ of the
  muon. \cite{Yao:2006px}} nor have there been any definitive indications of new
particle production at high energy colliders. This suggests the scale
of new physics is quite high (greater than 1 TeV). On the other hand,
within the context of the Standard Model, electroweak observables
require a light Higgs which, as discussed above,  is not easy to
reconcile with loop corrections to the Higgs mass if the new physics
resides above 1 TeV. In these models, it is often difficult to
introduce a large quartic coupling for the Higgs to raise its physical mass above the LEP limit, while still
protecting its mass term against corrections above the TeV scale.

Thus, the most attractive possibility is a BSM model in which the
Higgs bosons with large vevs (and hence large $ZZ$ and $WW$ couplings)
have mass of order $100\gev$ and new physics resides at scales
significantly below $1\tev$ while being consistent with current high
precision observables. In appropriately constructed BSM models, the
latter can be achieved.  However, in many models, having Higgs bosons
below $100\gev$ leads to an inconsistency with the limits from LEP
searches for Higgs bosons.  Consider first the Standard Model where
there is only one Higgs field and one Higgs boson. LEP has placed a
limit on the SM Higgs boson, $\hsm$, of $\mhsm>114.4\gev$.  Except in
a few non-generic corners of parameter space, this limit also applies
to the lightest CP-even Higgs boson of the Minimal Supersymmetric
Model (MSSM) and of many other BSM theories.  To escape this limit
there are basically two possibilities: (i) pushing the Higgs mass to
higher values where fine-tuning of parameters becomes an issue or
(ii) constructing models where the Higgs bosons with large $ZZ$
coupling have mass at or below $100\gev$ but were not detected at LEP
by virtue of having nonstandard decays that existing LEP analyses are
not sufficiently sensitive to.  It is a survey of the unusual decay
possibilities that is the focus of this review.

We shall begin by reviewing the motivations for nonstandard Higgs
decays in Section \ref{sec:motivation}, discussing model-dependent and
model-independent motivations, with particular attention to
supersymmetry, especially the next-to-minimal supersymmetric Standard Model
(NMSSM). We review the broad set of existing LEP Higgs searches in section
\ref{sec:lepsearches}. There, we emphasize that Higgs bosons of low mass
avoid the
normal LEP limits only if the primary Higgs decays are 
into non-Standard-Model particles each of which in turn decays to
SM particles.  Such decays are termed 
cascade decays.
In Section \ref{sec:naturalsusy}, we specialize to the
motivation for nonstandard Higgses from natural electroweak symmetry breaking in supersymmetry theories.   
In Section \ref{sec:higgslink}, we discuss generally the possibilities of extended Higgs sectors;  
then we focus on the best studied cases in the
context of the NMSSM in Section \ref{sec:nmssm}.  In
Section \ref{sec:bfactories}, we discuss some implications of nonstandard
Higgs physics for $B$-factories.  The LHC implications are discussed
in Section \ref{sec:lhc}.  Finally, in Section \ref{sec:conclusion} we
conclude.

\section{MOTIVATION FOR NONSTANDARD HIGGS BOSON DECAYS}
\label{sec:motivation}
The motivation for nonstandard Higgs boson decays comes from two
sources. First, a wide variety of theories predict new, neutral
states, affording the Higgs new channels into which it can decay. This
common feature motivates us to consider such decays irrespective of
anything else. However, naturalness can also be a significant
motivation. A theory that is ``natural'' is one in which the correct $Z$ boson
mass is obtained without any significant fine-tuning of the
fundamental parameters of the model: for example, the GUT-scale
soft SUSY breaking parameters in supersymmetric models.

There is often a tension in theories beyond the Standard Model between
naturalness and achieving a Higgs boson mass above the LEP limit.
This has been especially well studied within the context of
supersymmetry. By allowing the Higgs to decay into new final states,
one can have a lighter Higgs, and a more natural theory.  Indeed, the
recent interest in nonstandard Higgs decays was spurred by the
observation that the tension between natural electroweak symmetry
breaking in supersymmetric models and not seeing the Higgs boson at
LEP can be completely eliminated in models in which $h \to b \bar b$
is not the dominant decay mode of the SM-like Higgs
boson~\cite{Dermisek:2005ar}.

In the SM, there is just one Higgs boson and its dominant decay mode
is $\hsm \to b \bar b$ when $\mhsm\lsim 140\gev$. Although we have not yet made a definitive observation of this new state, a wide variety of tests at high energy experiments have already constrained its properties.
In particular, precision electroweak tests have continually suggested that the Higgs boson is light and accessible at LEP2.  The latest fits give an upper bound of 144 GeV at 95\% CL with a central value of 76 GeV \cite{lepewwg}.  Compared to the direct search bound of 114.4 GeV, it can be seen that there is some mild tension between these two Higgs bounds.  However, as \cite{Chanowitz:2002cd} points out, the story is  more complicated.  Notably, the measured forward-back asymmetry for $b$ quarks ($A^b_{FB}$) favors a heavy Higgs, but also is the most discrepant with the Standard Model fit (with a pull of about 3$\sigma$).  
Taking into account most of the data \cite{Chanowitz:2002cd}, the Standard Model electroweak fit has a poor confidence level of 0.01, whereas leaving out the most discrepant measurements improves the fit to a CL of 0.65.  However, then the best fit Higgs mass is 43 GeV, making the indirect Higgs bound more strongly in disagreement with the direct search limit.    

New measurements within the Standard Model have continued to support
this preference for a light Higgs.  In particular, the precise top and
$W$ mass measurements from Tevatron Run II have both gone in this
direction.  The constraint of the top and $W$ mass on the Higgs mass is well known, see e.g. \cite{Ferroglia:2004jx}. 
As of right now, the precision electroweak fit is
inconsistent with the direct search limit at the 68\% CL.  This
includes fitting $A^b_{FB}$, so excluding that measurement would
increase the discrepancy between the two limits.  Thus, even without
specifying a particular theory of physics, we see there is some
tension for the Standard Model Higgs at present, and this motivates us
to consider what possibilities exist for a light Higgs, and in
particular, one lighter than the nominal SM limit from LEP.

Furthermore, there were interesting excesses at LEP2 in Higgs
searches, which suggest there could be nonstandard Higgs physics.  The
largest excess ($2.3 \sigma$) of Higgs-like events at LEP was in the
$b\bar b$ final state for a reconstructed mass $M_{b\bar b}\sim 98 \gev$~\cite{Barate:2003sz}.
The number of excess events is roughly 10\% of the number of events
expected from the Standard Model with a 98 GeV Higgs boson.  Thus,
this excess cannot be interpreted as the Higgs of the Standard Model
or the SM-like Higgs of the MSSM.\footnote{In the MSSM, this excess can
  be explained by the lighter CP-even Higgs has highly reduced coupling to $ZZ$; see
  \eg\ \cite{Drees:2005jg}. This
  explanation doesn't remove the fine-tuning problem since it is the
  heavy CP-even Higgs which is SM-like and has to satisfy the 114 GeV limit.
  For a detailed discussion and references, see
  Ref.~\cite{Dermisek:2007ah}.}  However, this excess is a perfect
match to the idea of nonstandard Higgs decays, since the nonstandard
decay width reduces the branching ratio to Standard Model modes.

In the Standard Model, the Higgs has strong $O(1)$ couplings to the
$W, Z$ and top quark, but quite weak couplings to other fermions.
This means that for a Higgs mass that is below threshold for on-shell
$WW$ decays, the decay width into standard modes (in particular,
$b\bar b$) is quite suppressed.  A Higgs of mass, \eg, 100 GeV has a
decay width into Standard Model particles that is only 2.6 MeV, or
about $10^{-5}$ of its mass.  Consequently, the branching ratios to SM
particles of such a light Higgs are easily altered by the presence of
nonstandard decays; it doesn't take a large Higgs coupling to some
new particles for the decay width to these new particles to dominate
over the decay width to SM particles; the earliest studies pointing
this out of which we are aware are \cite{Gunion:1984yn,Li:1985hy,Gunion:1986nh}.

As one, but perhaps the most, relevant example, let us consider a
light Higgs with SM-like $b\anti b$ coupling and compare the decay
width $h\to b\anti b$ to that for $h\to aa$, where $a$ is a light
pseudoscalar Higgs boson.  Writing $\call\ni g_{h aa}haa$ with $g_{haa}=c\, {g m_h^2\over 2\mw}$
 and ignoring phase space suppression, we find 
\bea 
{\Gamma(h\to
  aa)\over \Gamma(h\to b\anti b)}&\sim & 310\, c^2\left({m_{h}\over
    100\gev}\right)^2.
\eea
This expression includes QCD corrections to the $b\anti b$ width as given in HDECAY \cite{Djouadi:1997yw}; these are evaluated for a 100 GeV Higgs and decrease the leading order $\Gamma(h\to b \anti b)$ by about 50\%.  
The decay widths are comparable for $c\sim 0.057$ when $m_h=100\gev$. Values of $c$
at this level or substantially higher (even $c=1$ is possible) are generic in BSM models
containing an extended Higgs sector. Further, both the
$h\to aa$ and $h\to WW$ decays widths grow as $m_h^3$, so that,
assuming SM $hWW$ coupling, $\Gamma(h\to aa)=\half c^2 \Gamma(h\to
WW)$ when neither is kinematically suppressed.

From a theoretical perspective, many BSM theories
have light neutral states such as nonstandard Higgs bosons, axions, neutralinos, sneutrinos,
\etc\ which are difficult to detect directly at existing
colliders. Typically, the main constraint on such light neutral states arises
if they contribute to the invisible $Z$ width. Thus, there are 
no strong constraints on their masses as long as their
coupling to the $Z$ is suppressed.  As a result, many of the light
states in BSM models can be light enough that a pair of them  
may appear in the decays of the 
Higgs boson. And, as discussed above, even a weak coupling of the
Higgs boson to these light BSM particles can cause this nonstandard
decay to dominate over the standard decay width.  

In many cases, the LEP2 constraint on the mass of the Higgs boson is
much weaker in the resulting final state than is the case if the Higgs
boson decays to either (a) a purely invisible final state or (b) a
final state containing just a pair of SM particles; for either final state,
the LEP2 data requires that the Higgs mass be greater than $114\gev$
if the Higgs $ZZ$ coupling is SM-like. This is because these two final
states are avoided if the light states are unstable, resulting in a
high multiplicity final state cascade decay with some visible
particles.  Note that since the cascade is initiated by Higgs decay to
just a pair of nonstandard particles, there is no additional phase
space suppression relative to a pair of SM particles and the
nonstandard pair can easily dominate despite the ultimate final state
containing many particles.  The importance of cascade decays
particularly emerged in early studies of the MSSM
\cite{Gunion:1984yn,Gunion:1986nh,Gunion:1988yc}, $E(6)$ models
\cite{Gunion:1987jd} and the NMSSM \cite{Ellis:1988er}.  These and
other models (such as triplet Higgs models and left-right symmetric
models) with cascade decays of one Higgs boson to a pair of lighter
Higgs bosons or supersymmetric particles were summarized in the Higgs
Hunters Guide \cite{Gunion:1989we,Gunion:1992hs}, which contains 
references  to the original work.

In more extreme models, LEP2 constraints are ineffective for even quite
light Higgs bosons. One particular example is the early work of
\cite{Espinosa:1998xj} in which there are many Higgs fields that 
mix with one another and share
the SM Higgs field vev. In this case, the physical Higgs eigenstates
also share the $ZZ$-Higgs coupling.
If the Higgs eigenstates are also spread out in mass, perhaps slightly
overlapping within relevant experimental resolutions (the worst case),
they could easily have avoided detection at LEP2
even if they have mass significantly below $100\gev$ and 
decay to a pair of SM particles. In fact, however,
such models typically have at least modest triple-Higgs
couplings and thus many of these multiple Higgs bosons would decay
primarily to a pair of lighter Higgs bosons each of which might then
decay either to a pair of SM particles or perhaps to a pair of still
lighter Higgs bosons.  A related model is that of 
\cite{Binoth:1996au} in which many unmixed (and, therefore, stable)
Higgs singlet fields are present and couple strongly to the SM Higgs 
field.  The SM Higgs will then decay primarily to pairs of singlet
Higgs bosons, yielding a very large SM Higgs width for the invisible 
final states. Because of the large width, the corresponding signal would
have been missed at LEP2 so long as the SM Higgs does not have mass
too much below $100\gev$.

To summarize, there is a wide open window for Higgs decays to light
unstable states with small coupling to the $Z$.  Indeed, beyond the
Standard Model theories having a mass light compared to the $WW$ threshold for
Higgs bosons that couple strongly to $WW, ZZ$ will generically
have nonstandard Higgs phenomenology.  Light states are ubiquitous in
BSM theories and could potentially be light enough for the Higgs to
decay into a pair of them.  Given that the decay width to a pair of SM
particles is so small for such Higgs bosons, the decay into a pair of
BSM states can easily dominate even when the relevant coupling is not
particularly strong.  Thus, the Higgs bosons associated with the Higgs
fields that give mass to the $W$ and $Z$ are highly susceptible to
having nonstandard Higgs phenomenology.  This will be illustrated in
greater depth as we discuss some particularly attractive model
realizations of such decays.

\section{LEP SEARCHES FOR THE HIGGS}
\label{sec:lepsearches}

Clearly, it is crucial to understand whether decays of a Higgs
boson to non-SM particles allow consistency with existing LEP
limits when the Higgs has mass below $100\gev$.  Although much
attention is focused on the SM Higgs search at LEP, there are actually
a wide variety of searches which were performed, constraining many
scenarios of nonstandard Higgs decays for light ($\lsim$ 114.4 GeV)
Higgses. We summarize here these constraints.

The dedicated Higgs searches at LEP2 encompass an impressive array of
possible Higgs decay topologies.  Since the Higgs is dominantly
produced in association with a $Z$ boson, the search topology
generally involves both the Higgs and $Z$ decay.  The searches give a
constraint on the product \begin{eqnarray} \xi_{h\to X}^2 \equiv
  \frac{\sigma(\epem \to Z h)}{\sigma(\epem \to Z h)_{SM}}\;
  Br(h \to X).
\end{eqnarray} 
The cross section $\sigma$ for Higgs production scales as the coupling
$g_{ZZh}$ squared, so the first factor is equivalently the square of
the ratio of this Higgs' coupling to the Standard Model value.  In Section~\ref{sec:motivation},
it was argued that precision electroweak results suggest that the nonstandard Higgs
has nearly standard couplings to Standard Model particles, so this
factor is close to one.~\footnote{In some cases, the SM $ZZ$
  coupling squared is shared among several Higgs bosons.  This is not
  typically the case for generic parameter choices for BSM models.
  Thus, in this section, when we refer to ``the Higgs'', we will be
  presuming that the Higgs has SM-like $ZZ$ and $WW$ couplings, but
  will allow for the possibility of nonstandard decays.}  The second
factor, $Br(h \to X)$, is the branching ratio of the Higgs decay in
question.  Now, we will discuss the relevant LEP2 Higgs searches for
our purposes.  Most stated mass limits are the 95\% CL lower bounds
assuming that $\xi_{h\to X}^2 =1$.

\vskip 0.15in
\noindent {\it Standard Model Higgs:} For any Higgs that is SM-like in
its couplings {\it and} decays, LEP limits are strongest for the dominant
Higgs decays into $b\bar{b}, \tau \bar{\tau}$.  LEP combined limits on
the SM Higgs~\cite{Barate:2003sz} require $\mh \geq 114.4$ GeV.
This study also includes the strongest limits on $h \to b\bar{b},\tau
\bar{\tau}$ rates with limits of about 115 GeV if the decay is
exclusively into either decay mode.

\vskip 0.15in
\noindent {\it Two parton hadronic states (aka Flavor-Independent):} 
In this analysis, the two parton decays of a SM-like Higgs were constrained. 
The analyses use the two parton final state that was least sensitive
to the candidate Higgs mass and details of the $Z$ decay.  The strongest
constraint is the preliminary LEP-wide analysis \cite{:2001yb}
requiring $m_h \geq 113$ GeV.

\vskip 0.15in
\noindent {\it Gauge Boson Decays (aka Fermiophobic):} 
This analysis focuses on two gauge boson decays of the Higgs, usually
assuming that the Higgs coupling to SM fermions is suppressed.  The
final states that are considered are $WW^*$, $ZZ^*$ as well as
photons. Assuming SM-like coupling to $ZZ^*$ and $WW^*$, implying the SM
decay width into gauge bosons, there is a limit of $m_h \geq 109.7$
GeV, while decays exclusively to two photons have a limit of $m_h \geq
117$ GeV \cite{Rosca:2002me}.

\vskip 0.15in
\noindent {\it Invisible Decays:} 
In this analysis, the Higgs is assumed to have SM-like $ZZ$ coupling
but to decay with 100\% branching into stable neutral
noninteracting particles.  The most stringent constraints are from an
older preliminary LEP-wide analysis with a limit $m_h \geq 114$ GeV,
see
\cite{unknown:2001xz}.
Since this constraint is so strong, the implication is that 
a nonstandard Higgs  must decay primarily 
into a state containing at least some visible particles if it is to have mass below $114\gev$.

\vskip 0.15in
\noindent {\it Cascade Decays:}
These constraints are relevant for the important nonstandard Higgs
decay where the Higgs decays into two secondary particles, such as a
pair of scalars $\phi$, and those scalars decay into $Y$ (i.e. $h \to
2 \phi \to 2Y$).  OPAL \cite{Abbiendi:2004ww} and DELPHI
\cite{Abdallah:2004wy} looked at $b$ decays ($Y \equiv b\bar{b}$), while
a LEP-wide analysis \cite{Schael:2006cr} has constrained both $b$ and $\tau$
decays.
For $h\rightarrow 2\phi \rightarrow 4b$ the limits are 110 GeV for
a Higgs produced with SM strength.  For other intermediate scalar
decays, $\phi \to 2g, c\bar{c}, \tau\bar{\tau}$, the best model-independent 
exclusions are from OPAL's analysis when the mass of the
scalar is below $b\bar{b}$ threshold. These limits are given in
\cite{Abbiendi:2002in}. It will be very important to note that this
latter analysis is restricted to Higgs masses in the range $45-86$
GeV.

\vskip 0.15in
\noindent {\it Model-Independent Decays:} 
This is the most conservative limit on the Higgs boson.  It assumes
that the Higgs is produced with a $Z$ boson and looks for electron
and muon pairs that reconstruct to a $Z$ mass, while the Higgs decay
process is unconstrained.  This study was done by
OPAL, giving a limit of $m_h \geq 82$ GeV
\cite{Abbiendi:2002qp}. 

Notice that these Higgs searches essentially exclude all Higgs decays
into a {\it pair} of Standard Model particles of a single Higgs with
SM-like $ZZ$ coupling and mass below about $113\gev$.  The only
possibilities not mentioned above are Higgs decays into a pair of electrons
or muons.  However, even though there is no dedicated search of this
type, such Higgs decays would give a large enhancement to charged lepton
events at LEP2.  Since $WW$ and
$Z e^+ e^-$ production at LEP2 was accurately measured to be consistent with the Standard Model
\cite{Alcaraz:2006mx}, limits on such decays would presumably be near the kinematic
limit.  Thus, it is essentially impossible for any Higgs boson (with SM-like $ZZ$
coupling) to have mass much below $114\gev$ if it decays entirely 
into any single mode or combination of modes, each of which contains just a pair of SM particles.

However, this does not rule out Higgs decays into a higher
multiplicity state.  Take for instance the cascade decay of the Higgs
into $4b$.  This has a weaker constraint than the $2b$ search,
although it only lowers the limit on the Higgs mass to 110 GeV.  More
drastically, the decay of the Higgs into $4\tau$ allows a Higgs as
light at 86 GeV.~\footnote{OPAL's examination of the $4\tau$ decay cut
  off the analysis at 86 GeV, so it is not clear what the reach of LEP
  is for a Higgs decaying in predominantly to $4\tau$.  However, it
  could reasonably be in the 90-100 GeV range.}  This is a substantial
weakening of the limit as compared to the di-tau search limit of 115
GeV.  Having said this, it does not mean that general high
multiplicity decays are completely safe.  First of all, the model
independent decay search requires that the Higgs be heavier than 82
GeV.  Dedicated LEP2 searches to particular decay topologies would of
course be more stringent than this.  However, absent these dedicated
limits, we will give plausible arguments that certain nonstandard
decays allow lighter Higgses. Our arguments will be based on applying existing LEP2 analyses to these
scenarios.  The estimated limits on such Higgs
decays obtained in this way should only be taken as a guideline.

\subsection{Decay Topologies Consistent With LEP Searches}
Before we go into more specific details of specific BSM models, let us
review the possibly important topologies and decays which are consistent with
the existing data. At the present, there are no strong constraints on
decays for Higgses above the LEP kinematical limit, and we shall
return to those cases later. We shall begin by focusing on what
topologies are allowed in light of the existing search data.

As we have already made clear, {\em decays of the Higgs into two body
  SM states are essentially as constrained as the SM Higgs}. This
compels us to consider decays into new states. If these states were
neutral, stable and weakly interacting, they would contribute to the
invisible Higgs search. Thus, to evade the strongest LEP limits, the
Higgs  must decay to a final state containing at
least one unstable particle which does not decay invisibly.

The simplest possible decay process is, $h \rightarrow 2a \rightarrow
4x$, where $x$ is some SM state. $x=b$ is already very constrained,
but $x=\tau$ is not constrained if $m_h>86\gev$ \cite{Schael:2006cr}, and no
explicit limits have been placed on situations where $x$ is a light,
unflavored jet for $a$ masses above 10 GeV, although one
can reasonably extrapolate limits in the range of 90 GeV from other
analyses \cite{Chang:2005ht}. Naively, it is hard to imagine that
cases where $x=e,\mu,\gamma$ are not excluded up to nearly the LEP
kinematical limit; however, no analyses have been explicitly 
performed and the LEP collaborations are not prepared to make an
explicit statement. 

More complicated decay topologies can arise when there are multiple
states below the Higgs mass, for instance a bino and a singlino (which
appear in generalized supersymmetric models). Assuming $R$-parity
conservation, such decays are typically characterized by two Standard Model fermions and missing energy. For instance, a particularly plausible
decay mode is $h \rightarrow( \wtil\chi_1 )\wtil\chi_0\rightarrow (\wtil\chi_0 f
\bar f) \wtil\chi_0$, where the $\wtil \chi_0$ is the LSP.  
Typically, a single $f \bar f$ mode does not dominate, as the decay
often includes multiple off-shell sleptons or an off-shell $Z$-boson.
As a result, for plausible branching ratios, such decays are allowed for
Higgs masses in the range $90-100$ GeV \cite{Chang:2007de}.

Most of the above mentioned nonstandard decays arise in the NMSSM and in
closely related variants of it. Moreover, the NMSSM is well studied
and is the simplest supersymmetric model in which it has been
explicitly shown that fine-tuning and naturalness problems can be
eliminated when the Higgs boson has mass at or below $\sim 100\gev$.
Thus, in the next section we turn to a detailed discussion of
naturalness in supersymmetric models and in the MSSM in particular.
In Section~\ref{sec:nmssm}, we will explain how fine-tuning can be
absent in the NMSSM by virtue of nonstandard Higgs decays allowing
$\mh\sim 100\gev$ and then consider the crucial new Higgs signals
within the NMSSM as well as its generalizations.

\section{NATURAL ELECTROWEAK SYMMETRY BREAKING IN SUPERSYMMETRY}
\label{sec:naturalsusy}

Within supersymmetry, the importance of minimizing fine-tuning provides a
particularly strong motivation for nonstandard Higgs decays.
Consider first the MSSM.  It
contains two Higgs doublet fields, $H_u$ and $H_d$, (with coupling to
up type quarks and down type quarks/leptons, respectively). EWSB
results in five physical Higgs states: light and heavy CP-even Higgs
bosons, $h$ and $H$, the CP-odd Higgs boson $A$, and charged Higgs
bosons $H^\pm$.  One typically finds that the $h$ is SM-like in its
couplings to gauge bosons and fermions.  Further, since the model
predicts $m_h<140\gev$, $h \to b \bar b$ is then the dominant decay
mode.~\footnote{For discussion of the possibility that $H$ is SM-like
  or that $h$ and $H$ share the coupling to $ZZ$ and $WW$ see {\it
    e.g.}~\cite{Dermisek:2007ah} and references there in. Fine-tuning
  can be ameliorated but not eliminated in such models. } As a
result, the LEP limit of $\mh>114.4\gev$ applies and, as discussed below,
will, in turn, imply a significant fine-tuning problem.  It
is only by turning to more general supersymmetric models that
fine-tuning can be avoided. In more general models such as the NMSSM,
the SM-like nature of the lightest CP-even Higgs boson remains a
generic feature, but it is not necessarily the case that it dominantly
decays to $b \bar b$. As we have discussed, due to the small size of
the SM-like $h b \bar b$ coupling, any new decay modes that are
kinematically accessible tend to dominate the Higgs
decays and can allow $\mh\leq 100\gev$ to be consistent with LEP
limits. For such $\mh$ values, fine-tuning problems can be absent.
Phenomenological implications of such a scenario are dramatic. In
particular, prospects for detecting the $h$ at the Tevatron and the
LHC are greatly modified.

 
Let us now focus on the issues of naturalness and fine-tuning.
For the triggering of EWSB by SUSY breaking to be natural,
the superpartners must be near the EW scale. This is because 
the mass of the $Z$ boson, determined by
minimizing the Higgs potential, is related to the supersymmetric
Higgs mass parameter $\mu$ and the soft SUSY breaking mass squared 
parameter for $H_u$, for $\tan \beta \ge 5$, by: 
\be \frac
{M_Z^2}{2} 
\simeq  -\mu^2 (M_Z) - m_{H_u}^2 (M_Z). 
\label{eq:MZ} 
\ee 
The EW scale value of $m_{H_u}^2$ depends on the boundary conditions of 
all soft SUSY breaking parameters through renormalization group (RG) evolution. 
 For a given $\tan \beta$, we can solve the 
RG equations exactly and express the EW values of $m_{H_u}^2$,
$\mu^2$, and consequently $M_Z^2$ given by Eq.~(\ref{eq:MZ}), in terms
of all GUT-scale parameters; although we consider the GUT-scale as an example, the conclusions 
do not depend on this choice. For $\tan \beta =10$, we
have:
\bea M_Z^2 & \simeq & -1.9 \mu^2 + 5.9 M_3^2 -1.2 m_{H_u}^2 + 1.5 m_{\tilde{t}}^2
  - 0.8 A_t M_3 + 0.2 A_t^2 + \cdots,
\label{eq:MZ_gut} 
\eea 
where parameters appearing on the right-hand side are the GUT-scale
parameters. Here, $M_3$ is the
$SU(3)$ gaugino mass, $A_t$ is the trilinear stop soft SUSY breaking mixing
parameter and for simplicity we have defined $m^2_{\tilde{t}} \equiv \half(
m_{\tilde{t}_L}^2 + m_{\tilde{t}_R}^2)$, the latter being the
soft SUSY breaking stop mass squared parameters.  Other scalar masses
and the $U(1)_Y$ and $SU(2)$ gaugino masses, $M_1$ and $M_2$, appear
with negligible coefficients and we neglect them in our discussion.
The coefficients in this expression depend weakly on $\tan \beta$ 
and on $\log (\mgut/\mz)$.  We can express the EW scale
values of the stop mass squared, gluino mass and top trilinear
coupling in a similar way; for $\tan \beta = 10$ we have:
\bea
m_{\tilde{t}}^2 (M_Z) & \simeq & 5.0 M_3^2 + 0.6 m_{\tilde{t}}^2   + 0.2 A_t M_3 \label{eq:mstop_gut} \\
M_3 (M_Z) & \simeq & 3 M_3 \label{eq:M3_gut} \\
A_t (M_Z) & \simeq & - 2.3 M_3 + 0.2 A_t. \label{eq:At_gut} 
\eea
From Eqs.~(\ref{eq:MZ_gut}), (\ref{eq:mstop_gut}) and
(\ref{eq:M3_gut}), we see the usual expectation from SUSY, 
\be
M_Z \simeq m_{\tilde{t}_{1,2}} \simeq m_{\tilde{g}},
\label{eq:expectation}
\ee 
when all the soft SUSY breaking
parameters are comparable. Furthermore, neglecting
terms proportional to $A_t$ in Eqs.~(\ref{eq:At_gut}) and
(\ref{eq:mstop_gut}) we find that a typical stop mixing is 
$A_t(\mz)/m_{\tilde{t}}(M_Z) \lsim 1.0$.  This result has an important implication for the Higgs mass.

The mass of the $h$ is approximately given as: 
\be
m_h^2 \simeq M_Z^2 \cos^2 2\beta + \frac{3G_F m_t^4}{\sqrt{2} \pi^2}
\left\{ \log \frac{m_{\tilde{t}}^2(\mz)}{m_t^2} +
  \frac{A_t^2(\mz)}{m_{\tilde{t}}^2(\mz)} \left(1-\frac{A_t^2(\mz)}{12 m_{\tilde{t}}^2(\mz)} \right)
\right\}. \label{eq:mh_mix} 
\ee 
where the first term is the tree level result and the second term is
the dominant one loop correction \cite{Okada:1990vk, Haber:1990aw,
  Ellis:1990nz, Ellis:1991zd}.  At tree level, $m_h \leq M_Z \simeq
91$ GeV. It can be increased beyond this value either by increasing
the mixing in the stop sector, $A_t(\mz)/m_{\tilde{t}}(\mz)$, or by
increasing the stop mass, $m_{\tilde{t}}(\mz)$. As we have learned,
the typical mixing in the stop sector achieved as a result of RG
evolution from a large range of high scale boundary conditions is
$A_t(\mz)/m_{\tilde{t}}(M_Z) \lsim 1.0$.  With this typical mixing, we
obtain the typical Higgs mass, $m_h \simeq 100$ GeV. In order to push
the Higgs mass above the LEP limit, 114.4 GeV, assuming the typical
mixing, the stop masses have to be $\gsim 1\tev$.~\footnote{The Higgs
  mass is maximized for $|A_t(\mz)/m_{\tilde{t}}(\mz)| \simeq 2$,
  which corresponds to the maximal mixing scenario.  In this case,
  $m_{\tilde{t}}(\mz)$ can be as small as $\sim$ 300 GeV without
  violating the bound on $m_h$ from LEP. However it is not trivial to
  achieve the maximal mixing scenario in models. For more details see,
  \eg, the discussion in Refs.~\cite{Dermisek:2006ey,
    Dermisek:2007yt}. }~\footnote{In models beyond the MSSM, with
  extended Higgs sectors or extended gauge symmetries, the tree level
  prediction for the Higgs mass can be increased, see
  Refs.~\cite{Brignole:2003cm,Dine:2007xi} and references therein. This
  increase is not automatic and typically requires nontrivial
  assumptions.  For example, in the NMSSM (as defined in Sec. 3)
  assuming perturbativity up to the GUT scale, the tree level
  prediction for the Higgs mass can be increased only by a small
  amount.}

The need for 1 TeV stops is in direct contradiction with the usual
expectation from SUSY, Eq.~(\ref{eq:expectation}). The hierarchy
between the scale where SUSY is expected and the scale to which it is
pushed by the limit on the Higgs mass requires a precise
cancellation, at better than 1\% precision, between the
soft SUSY breaking terms and the $\mu$ term appearing on the
right-hand side of Eq.~(\ref{eq:MZ_gut}) in order to recover the
correct value of the $Z$ mass.  This is the explicit realization
of the fine-tuning problem in the MSSM and, as described in the
introduction, is closely related to the little hierarchy problem.

The solution to the fine-tuning problem in models in which the SM-like
Higgs decays dominantly to non-SM particles is straightforward.  If
the $h \to b \bar b$ decay mode is not dominant the Higgs boson does
not need to be heavier than 114 GeV, it can be as light as the typical
Higgs mass or even lighter depending on the experimental limits placed on the
dominant decay mode. If the strongest limit is $\leq 100\gev$, there is no need
for large superpartner masses and superpartners can be as light as
current experimental limits allow. In Section~\ref{sec:lepsearches},
we reviewed the experimental limits on the mass of the SM-like Higgs
boson in various decay modes. Quite surprisingly, only if the Higgs
decays primarily to two or four bottom quarks, two jets, two taus or
to an invisible channel (such as two stable LSPs), is the LEP limit
on $m_h$ above $100\gev$. Most other decay topologies have not been studied directly,
and applications of other searches (e.g., the sensitivity of the flavor-independent two jet search to the general four jet topology)
typically imply weak limits.~\footnote{We are assuming that the Higgs
  is produced with standard strength. A four bottom quark decay for a
  Higgs which is strongly mixed as in \cite{Barbieri:2007tu} may be
  allowed for Higgs mass below $100\gev$ within the existing
  constraints. Even so, a state lighter than $\sim 105$ GeV which
  decays to four bottom quarks cannot have a coupling larger than $40
  \%$ of standard-model strength. Such models represent a different approach than we principally consider here.} Moreover, it is reasonable to take a very conservative approach in which one does not extrapolate LEP limits beyond their explicit analyzed
topology. 
 
 Regardless, LEP limits on $m_h$ for other decay
modes are generally below $90\gev$ and would therefore not place a constraint on
superpartner masses.  Since $\mh\sim 90-100\gev$ is the generic
prediction for supersymmetric models in which there is no fine
tuning, those supersymmetric models where these alternate decay modes
are dominant automatically provide a solution to the fine-tuning
problem.~\footnote{In specific models, avoiding the fine-tuning
  problem might require another tuning of parameters in order to make
  an alternative decay mode for the Higgs boson dominant, see, \eg,
  Ref.~\cite{Schuster:2005py, Dermisek:2006wr} for the discussion of
  these issues in the NMSSM.}

Besides alleviating or completely removing the fine-tuning problem the
possibility of modified Higgs decays is independently supported
experimentally. As mentioned in Section~\ref{sec:motivation}, the largest Higgs excess suggested a nonstandard Higgs of mass 98 GeV, that only decayed 10\% of the time to Standard Model decay modes.
As we have discussed, from natural EWSB we expect the
SM-like $h$ to have mass very near 100 GeV, and this is possible in
any model where the SM-like Higgs boson decays mainly in a mode for
which the LEP limits on $m_h$ are below $100\gev$,
such as those mentioned earlier for which LEP limits run out at
$90\gev$.~\footnote{Another possibility would be that a weakly mixed
  state existed at 98 GeV but the dominant state coupling to the $Z$
  was above the LEP bound. (See,  \eg, \cite{Demir:2005kg}.)} The $h \to b \bar
b$ decay mode will still be present, but with reduced branching ratio.
Any $\br(h \to b \bar b) \lsim 30\%$ is consistent with experimental
limits for $m_h \sim 100$ GeV. Further, $\br(h \to b \bar b) \sim
10\%$ with $m_h\sim 100\gev$ provides a perfect explanation of the
excess.  This interpretation of the excess was first made in the NMSSM
with the $h \to aa \to \tau^+ \tau^- \tau^+ \tau^- $ mode being
dominant~\cite{Dermisek:2005gg}, but it clearly applies to a wide
variety of models.

Within the MSSM context, there is one scenario worth mentioning that
can alleviate fine-tuning. For example, for
$\tanb\sim 10$, $\mstopbar(\mz)\sim 300\gev$ and $A_t(\mz)\sim
-400\gev$, fine-tuning is moderate ($\sim 6\%$) and $\mh\sim
95-100\gev$, thereby providing a contribution to the $\sim 98\gev$ LEP
excess observed~\cite{Barate:2003sz} at LEP. 
In this case, LEP limits are evaded by virtue of substantial
Higgs mixing leading to greatly reduced $hZZ$ coupling; the $HZZ$
coupling is large but $\mhh$ is slightly above the LEP limit of
$114.4\gev$.~\footnote{Without such mixing, the $\mh>114\gev$ LEP
  limit applies and at least $3\%$ fine-tuning is necessary.}  In fact, for
$\tanb\sim 20$ one can simultaneously fit the $\sim 98\gev$ and $\sim
116\gev$ LEP excesses~\cite{Drees:2005jg,Demir:2005kg,Dermisek:2007ah}. These 
scenarios, however, require highly nongeneric boundary conditions at
the GUT scale and are clearly scenarios characterized by nearly
maximal mixing in the stop sector.  Another scenario is that in which
one allows large CP violation in the Higgs sector. Then, the physical
MSSM Higgs states will be mixtures of the CP-even $h$ and $H$ and the
CP-odd $A$; let us label the 3 resulting eigenstates as $H_{1,2,3}$.
It is possible to arrange $2m_{H_1}<m_{H_2}$ with the $H_2ZZ$ coupling
near maximal and $m_{H_2}\leq 100\gev$~\cite{Carena:2000ks}. This can be
consistent with LEP limits if the $H_2\to
H_1H_1$ cascade decay is dominant and $m_{H_2}<2m_b$.  Whether or not this scenario is fine-tuned has
not been studied, but one can speculate that the low mass of the $H_2$
would imply reduced fine-tuning. However, for $m_{H_1}$ sufficiently below the
Upsilon mass,  the rate for $\Upsilon\to \gamma H_1$ would typically
be large and inconsistent with limits \cite{Kreinick:2007gh} from $B$ factories. This is
because the $H_1$ is part of a doublet (unlike the NMSSM where
the $\ai$ is mainly singlet) and, 
since the $H_2$ is the SM-like Higgs, the $H_1$ (and $H_3$)
will have $\tanb$-enhanced coupling to $b\anti b$.

In summary, it is only models with nonstandard Higgs decays that can
completely avoid the fine-tuning problem.  They allow the Higgs boson
mass to be the $m_h\sim 100\gev$ value predicted from natural EWSB
while at the same time the now subdominant decay mode, $h \to b \bar b$,
with $\sim 10\%$ branching ratio can explain the largest excess of
Higgs-like events at LEP at $M_{b\bar b}\sim 98$ GeV. A SM-like $h$
with $m_h\sim 100\gev$ is also nicely consistent with precision
electroweak data.

\section{HIGGS AS A LINK TO NEW SECTORS}
\label{sec:higgslink}
The Higgs field plays a unique role in the Standard Model in that
$h^\dagger h$ is a complete Lorentz and gauge singlet, but is only
dimension two. As a consequence, it can couple to hidden sector scalar
fields (real or complex, the latter implying the presence of both
scalar and pseudoscalar mass eigenstates) through the renormalizable
operator $\phi^* \phi h^\dagger h$ or through the dimension three
trilinear interaction $\phi h^\dagger h + h.c.$. The latter can be
eliminated by requiring a symmetry under $\phi\to -\phi$. If we presume
that the trilinear term is absent then it is still the case that 
couplings of the type $h\phi\phi$ will be generated for the mass
eigenstates when the $h$ field acquires a vev. Similarly, the $h$ can
couple to (vector-like) SM singlet fermions through the dimension five
operator $\bar n n h^\dagger h$. Because the width of the Higgs
eigenstate (also denoted $h$) is small for $m_h<160\gev$, decays to a
pair of $\phi$'s or $n$'s can dominate the decay of the $h$, in the
former case with a perturbative dimensionless coupling and in the
latter case if the operator is suppressed by a scale near the weak
scale. We review here some of the possibilities arising from these
operators. Both of the above possibilities arise quite simply in the
NMSSM, but we first consider them in a general context.

There are many possible final signals in the case of $h\to \phi\phi$ decays.
The various cases depend upon whether $\phi$ acquires a vev or not and
on whether the $\phi$ couples to a new heavy BSM sector. If the $\phi$
field does
not acquire a vev and does not couple to some new BSM sector, then the
$\phi$ mass eigenstate
will be absolutely stable.  If the $h$ width remains narrow for such
decays, then LEP2 limits on an invisible $h$, requiring $m_h\geq
115\gev$ will apply.  However, this limit can be evaded if the
invisible width is very large.  Models~\footnote{We do not consider 
invisible decay modes
related to graviscalars and so forth that arise in theories with extra
dimensions.} in which this can happen include the case
of a large number of strongly coupled scalars 
\cite{Binoth:1996au} and, for some extreme parameter choices, 
the recent unparticle models (which can be
deconstructed in terms of a ``continuum'' of stable invisible
scalars, a subcase of the former class of models)~\cite{Delgado:2007dx}.  

If the purely singlet 
scalar couples to the mass of some heavy BSM fermions (\ie\ 
$(\lambda \phi +M)\bar \psi \psi$), then $h$ decays to four photons or
four jets can dominate the Higgs width
\cite{Dobrescu:2000jt,Chang:2006bw}. The four photon channel can also be
dominant in NMSSM models \cite{Arhrib:2006sx} where $h\to aa$ and the
$a$ is purely singlet and the $a\gamma\gamma$ coupling arises from
virtual loops containing supersymmetric gauginos and higgsinos. The
four photon mode would have been easily discoverable at LEP were it below
the kinematic bound while the four gluon decay would have been a challenge
\cite{Chang:2006bw}. If there are multiple states below the Higgs mass, then
very complicated decays, such as $h \rightarrow 6f$ or $h \rightarrow
8f$ can arise, where $f$ can represent various fermions.

A purely singlet scalar can in fact be very long lived if it decays
through loop-suppressed or non-renormalizable operators
\cite{Strassler:2006im,Strassler:2006ri,Chang:2006bw}. This allows
decays of the Higgs with significantly displaced vertices which may be
an intriguing avenue to search for new physics like that of ``Hidden Valleys''
\cite{Strassler:2006ri}.  However, such decays would likely have been
noticed had they been the dominant decay mode at LEP.

Let us now discuss the cases in which the
the singlet scalar field acquires a vev or there is an
$h^\dagger h \phi+h.c.$ component in the Lagrangian. In these cases,
the above discussion does not apply.
The light singlet state will typically mix with the non-singlet
Higgs boson(s) and thereby acquire a significant coupling to SM
particles, especially the light fermions. Thus, the cascade decays can arise, as described earlier, and reduce the fine-tuning.  For very light singlet states, most of the phenomenology is independent of the potential \cite{O'Connell:2006wi}, while at heavier masses there is more dependence. 

However, without considering specific decays, the mixing alone can influence fine-tuning in two separate ways. First, the Higgs mixing can push up the
mass of the heavy mass eigenstate, but at the cost of producing the
light state via Higgsstrahlung. At tree level, such mixing
does little to alleviate fine-tuning, but with loop effects included, the
reduction in fine-tuning can be significant \cite{Kim:2006mb}.  More
common is the reduction in fine-tuning due to the fact that a light
SM-like Higgs can decay to a pair of still lighter Higgses in such
models, thereby allowing the SM-like Higgs to have mass $\leq
100\gev$. As we shall discuss in Section~\ref{sec:lhc}, this scenario
will also make Higgs discovery at the LHC quite challenging.  As one
increases the number of states with which the $h$ field can mix
(including now both doublets and singlets in general), the primary
Higgs field (or fields) that couple to the $Z$ can be spread out among
many mass eigenstates.  This will make discovery difficult due to the
fact that such a model allows a multitude of Higgs to Higgs pair
decays, a reduction in the production rate for any one Higgs boson,
and an overlapping of the peaks for the individual states in any given
detection channel.  In particular, these effects can make LHC Higgs
detection essentially impossible
\cite{Espinosa:1998xj,Patt:2006fw,BahatTreidel:2006kx}. However, detection at a
future linear collider will be possible in the $\epem\to Z +X$ channel
provided sufficient integrated luminosity is available \cite{Espinosa:1998xj}.

If the singlet field acquires a vev and yet parameters are chosen so
that there is no mixing between the singlet $\phi$ particle state and the doublet
Higgs states, one must again consider whether or not the four photon and other highly
suppressed decay modes of the SM-like Higgs could be dominant. An
example is provided by the NMSSM. There,
aside from the loop induced $a\gamma\gamma$ coupling considered in
\cite{Arhrib:2006sx}, supersymmetric particle loops (for example, a
loop containing a $\wtil b$ and a gluino) also induce $a b\anti b$
couplings~\cite{Hodgkinson:2006yh}.  The latter can be up to one half
of SM-like strength (\ie\ like $\hsm b\anti b$ but with an extra
$\gamma_5$ in the coupling Lagrangian) for very high values of
$\tanb\sim 50$ and moderate superparticle masses.  This is more than
likely to swamp the loop-induced $a\gamma\gamma$ couplings in the
NMSSM. Generically, even for a purely singlet $a$, dominance of $a\to
\gamma\gamma$ over $a\to b\anti b$ will only be the case if $\tanb$ is
small. More generally, if the BSM sector, whose loops give rise to a
singlet-2photon coupling, contains any non-SM-singlet fields, one can
expect important singlet-$b\anti b$ couplings associated with loops of
the latter fields.

If the additional scalar is charged under some new gauge symmetry, a
strongly mixed Higgs can decay into new gauge bosons
\cite{Gopalakrishna:2008dv}. However, in order for this to dominate
the decay, the mass eigenstate must be significantly mixed ($\sin^2 \theta \sim 0.5$).

In the case of BSM fermions dominating the Higgs decay, Higgs decays to
right handed neutrinos are an interesting possibility \cite{Graesser:2007pc, Graesser:2007yj}. If left-handed neutrinos are
involved, decays to different fermions, i.e., $h \rightarrow \psi_1
\psi_2$ are possible \cite{deGouvea:2007uz,Chang:2007de}. Such decays with both
visible and missing energy are also capable of evading Higgs search
limits, as well as other new physics searches \cite{Chang:2007de}. Such states
could also be neutralinos in the NMSSM, in addition to neutrinos.

In the above discussion, we noted the generic possibility of coupling
the doublet Higgs structure $h^\dagger h$ to a SM singlet operator.
In fact, in supersymmetry such coupling has a very compelling
motivation as an extension of the MSSM.
The content of the MSSM in the matter and gauge sectors is fixed by
requiring a superpartner for each known particle of definite helicity.
In contrast, the choice of a two doublet Higgs sector is made purely
on the basis of minimality arguments (absence of anomalies and the
need to give mass to both up and down type quarks) 
and this choice gives rise to the
famous $\mu$-problem. Namely, phenomenology requires a term of the
form $\mu \what H_u\what H_d$~\footnote{Hatted fields denote
  superfields while unhatted fields are normal fields.} in the
superpotential with $\mu$ being a term with dimensions of mass with a
value somewhere between about $150\gev$ and $1\tev$, as opposed to the
natural values of $0$ or $\mgut$. Ideally, one would have no
dimensionful parameters in the superpotential, all dimensionful
parameters being confined to the soft SUSY breaking potential.

A particularly appealing extension of the MSSM which solves the $\mu$
problem is the introduction of a completely new sector of particles
which are singlets under the SM gauge symmetry.  As such, this extra ($E$)
sector would not spoil any of the virtues of the MSSM, including the
possibility of gauge coupling unification and matter particles fitting
into complete GUT multiplets. In addition, $E$-sector particles that
either do not mix or have small mixing with SM particles would have
easily escaped direct detection. Of course, if this $E$-sector is
completely decoupled from the SM then it plays no role in particle
physics phenomenology at accelerators.  Much more interesting is the
possibility that this sector couples to the MSSM through the Higgs
fields.  In particular, the $E$-sector can couple to the SM-singlet
$\what H_u \what H_d$ form appearing in the MSSM $\mu$ term in many
ways, including a renormalizable term (with dimensionless coupling) of
form $\lambda \what E \what H_u \what H_d$.  When the scalar component
of the singlet superfield $\what E$ acquires a vev, $\vev{E}=x$ (as a
result of SUSY breaking) an effective $\mu$ value, $\mueff=\lambda x$,
is generated. Such couplings would have a negligible effect on the
phenomenology involving SM matter particles, whereas they can
dramatically alter Higgs physics. In particular, the particle
couplings generated allow the 
lightest CP-even Higgs boson $h$ to decay into two of the particles
associated with the $E$-fields ($E$-particles) if the $E$-particles
are light enough, and these $h\to EE$ decays
can be dominant for even rather modest $\what E\what H_u\what H_d$ coupling
strength.

The implications for Higgs discovery follow some of the patterns
discussed above. In particular, when
$h$ decays to two lighter $E$-particles are dominant, the strategy for
Higgs discovery will depend on the way the $E$-particles appearing in
the decays of the $h$ themselves decay. The latter might decay predominantly
into other stable $E$-particles, in which case the MSSM-like $h$
decays mainly invisibly. More typically, however, the $E$-particles
mix with the SM particles via the couplings between the MSSM and
$E$-sector.  In particular, couplings between $E$-bosons and the MSSM
Higgs fields are generically present and imply that the Higgs mass
eigenstates are mixed.  In this case, the mostly $E$-particle light
Higgses will decay into $b \bar b$, $\tau^+ \tau^-$ or other quarks or
leptons depending on the model. Although $E$-particles would have
small direct production cross sections and it would be difficult to
detect them directly, their presence would be manifest through the
dominant Higgs decay modes being $h \to 4f$, where $4f$ symbolically
means four SM particles, \eg\ $b \bar b b \bar b$, $b\bar
b\tau^+\tau^-$, $\tau^+ \tau^- \tau^+ \tau^-$, $4 \gamma$ and so on.
The situation can be even more complicated if the $h$ decays to
$E$-particles that themselves decay into other $E$-particles which in
turn finally decay to SM particles. In such a case, the SM-like Higgs
would effectively decay into $8f$.  

Let us finally note that the presence of a singlet in the potential can lead to modifications in the early universe cosmology. In particular, it allows the possibility of a first-order phase transition \cite{Apreda:2001tj,Menon:2004wv,Profumo:2007wc}, which can arise consistent with LEP experiments.



\section{CASCADE DECAYS TO SCALARS IN THE NMSSM}
\label{sec:nmssm}
The cascade decay scenario described in Sections~\ref{sec:motivation},
\ref{sec:lepsearches} and \ref{sec:higgslink} already occurs in the
simplest extension of the MSSM, the next-to-minimal supersymmetric
Standard Model (NMSSM) which adds only one singlet chiral superfield,
$\widehat{S}$ to the MSSM. Phenomenologically similar scenarios arise
naturally in theories with additional $U(1)$'s
\cite{Han:2004yd,Barger:2006dh,Barger:2007im}.  The NMSSM particle
content differs from the MSSM by the addition of one CP-even and one
CP-odd state in the neutral Higgs sector (assuming CP conservation),
and one additional neutralino.  We will follow the conventions of
\cite{Ellwanger:2004xm}.  Apart from the usual quark and lepton Yukawa
couplings, the scale invariant superpotential is \vspace*{-.07in}
\beq \label{1.1} \lambda \ \widehat{S}
\widehat{H}_u \widehat{H}_d + \frac{\kappa}{3} \ \widehat{S}^3
\vspace*{-.11in} 
\eeq 
\noi depending on two dimensionless
couplings $\lambda$ and $\kappa$ beyond the MSSM.  
The associated trilinear soft terms are
\vspace*{-.1in}
\beq \label{1.2}
\lambda A_{\lambda} S H_u H_d + \frac{\kappa} {3} A_\kappa S^3 \,.
\vspace*{-.1in}
\eeq
The final two input parameters are
\vspace*{-.1in}
\beq \label{1.3} \tan \beta = h_u/h_d\,, \quad \mu_\mathrm{eff} = \lambda
s \,,
\vspace*{-.07in}
\eeq
where $h_u\equiv
\vev {H_u}$, $h_d\equiv \vev{H_d}$ and $s\equiv \vev S$.
These, along with $\mz$, can be viewed as
determining the three \susy\ breaking masses squared for $H_u$, $H_d$
and $S$ (denoted $\mhusq$, $\mhdsq$ and $\mssq$)
through the three minimization equations of the scalar potential.
Thus, as compared to the three
independent parameters needed in the
MSSM context (often chosen as $\mu$, $\tan \beta$ and $M_A$), the
Higgs sector of the NMSSM is described by the six parameters
\vspace*{-.1in}
\beq \label{6param}
\lambda\ , \ \kappa\ , \ A_{\lambda} \ , \ A_{\kappa}, \ \tan \beta\ ,
\ \mu_\mathrm{eff}\ .
\vspace*{-.1in}
  \eeq
We will choose sign conventions for the fields
such that $\lambda$ and $\tan\beta$ are positive, while $\kappa$,
$A_\lambda$, $A_{\kappa}$ and $\mu_{\mathrm{eff}}$ should be allowed
to have either sign.
In addition, values must be input for the gaugino masses
and for the soft terms related to the (third generation)
squarks and sleptons (especially $m_{\wtil t_L}^2$, $m_{\wtil t_R}^2$ and $A_t$) that contribute to the
radiative corrections in the Higgs sector and to the Higgs decay
widths. 

Of all the possible new phenomena the additional Higgses in the NMSSM
can lead to, perhaps the most intriguing one is the possibility of the
lightest CP-even Higgs decaying into a pair of the two lightest CP-odd
Higgses, $\hi\to\ai\ai$, where the latter are mostly
singlets~\cite{Gunion:1996fb,Dobrescu:2000jt,Dermisek:2005ar,Dermisek:2005gg,Dermisek:2007yt}.
Precisely this scenario can eliminate the fine-tuning of EWSB in the
NMSSM for $\mhi\sim 100\gev$~\cite{Dermisek:2005ar, Dermisek:2007yt}.
If $\br(\hi\to\ai\ai)>0.7$ and $\mai<2\mb$, the usual LEP limit on
the Higgs boson mass does not apply and the SUSY spectrum can be
arbitrarily light, perhaps just above the experimental bounds and 
certainly light enough for natural EWSB.  In addition, without any
further ingredients this scenario can completely explain the excess of
Higgs-like events in the $b\bar b$ channel at $M_{b\bar b} \simeq
98\gev$~\cite{Dermisek:2005gg}. Finally, the above $\ai$ scenario is
not itself fine-tuned. Starting from $\akap$ and $\alam$ values at the
GUT scale that are small (and therefore close to a $U(1)_R$ symmetry
limit of the potential), the RG equations yield $\akap$ and $\alam$
values at scale $M_Z$ that generically result in 
$\br(\hi\to\ai\ai)>0.7$ and $\mai<2\mb$ with some preference for
$\mai>2\mtau$ \cite{Dermisek:2006wr}. In addition, 
$M_Z$-scale values for the soft SUSY breaking parameters 
that correspond to there being no electroweak fine-tuning,
imply values of $\mhusq$, $\mhdsq$ and $\mssq$ at the GUT scale that
are all relatively small~\cite{Dermisek:2005gg}.  That is, the preferred GUT-scale boundary
conditions for the NMSSM are close to the no-scale type boundary
conditions where many of the soft SUSY breaking parameters are near
zero.

Higgs signals at colliders of all types are dramatically different in
the NMSSM models that have no fine-tuning.  One must look for $h\to
\ai\ai\to 4\tau$ or $4j$ (somewhat less preferred because of a need to tune
$\akap$ and $\alam$). We discuss collider implications in the
following sections.

\section{LIGHT HIGGSES AT {\boldmath $B$} FACTORIES}
\label{sec:bfactories}

We have seen that a particularly generic way in which a SM-like Higgs
with $m_h\sim 100\gev$ can escape LEP limits on the $h\to b\bar b$ and
$h\to b\bar b b\bar b$ channels, is for the $h$ to decay primarily to
two $E$-bosons which have mass below $2\mb$. The NMSSM scenario of
$\hi\to\ai\ai\to 4\tau$ is just one example of this generic
possibility. However, there is an interesting requirement within the
NMSSM scenario that we have not yet mentioned.  Namely,
$\br(\hi\to\ai\ai)$ is only large enough to escape LEP limits if the
$\ai$ is not purely singlet~\cite{Dermisek:2006wr}.  There must be
some mixing of the CP-odd singlet with the MSSM-like CP-odd Higgs that
is a residual from the two doublets.  Defining
\be
\ai\equiv \cta\, A_{MSSM}+\sta\, A_S\,,
\ee
where $A_{MSSM}$ is the MSSM-doublet CP-odd Higgs and $A_S$ is the
CP-odd (imaginary) component of the complex $S$ scalar field, one
finds (at $\tanb=10$ for example) that $|\cta|\gsim 0.06$ is required
for $\br(\hi\to\ai\ai)>0.7$.  The $\ai b\bar b$ coupling, given by
$\cta\tanb\, {m_b\over v}\,\bar b i\gamma_5 b\,\ai$, then has a lower bound
that is not so far from being SM-like in strength. Further, a light
$\ai$ with the required properties is most naturally obtained after RG
evolution of the relevant parameters from GUT-scale boundary
conditions when $|\cta|$ is near its lower bound, $|\cta|\sim 0.1$.

The lower bound on the $\ai b\bar b$ coupling has crucial consequences
at $B$ factories \cite{Dermisek:2006py,Fullana:2007uq}. In the NMSSM,
one finds~\cite{Dermisek:2006py} that for any given $\mai$ below
$M_{\Upsilon}$ (where $\Upsilon$ denotes the $1S$, $2S$ or $3S$ state)
there is a lower bound on $\br(\Upsilon\to \gamma \ai)$. Not
surprisingly, this lower bound is quite small. For example, to probe
$\mai$ values as high as $9.2\gev$ in $\Upsilon(1S)$ decays (such an
$\mai$ being within the preferred $\mai>2\mtau$ range but still
leaving some phase space for $\Upsilon(1S)\to \gamma \ai$ decay), one
needs to be sensitive at $B$ factories to $\br(\Upsilon\to \gamma
\ai\to \gamma \tau^+\tau^-)$ down to $\sim 10^{-7}$ for full coverage
of the possible scenarios.  Reaching this level is a challenge, but
not necessarily impossible using dedicated runs on one of the
$\Upsilon$ resonances. Search for non-universality (enhancement of the
$\tau^+\tau^-$ final state) in $\Upsilon$
decays to leptons without directly tagging the
photon may also be a useful approach~\cite{Fullana:2007uq}.

In more general $E$-sector scenarios, to have a SM-like $h$ with
$m_h\sim 100\gev$ again requires that there be one or more
$E$-bosons with mass below $2\mb$ and the cumulative branching ratio
for $h$ decay to these states must be $\gsim 0.7$. It is quite likely
that some of these states will have reasonable coupling to $b\bar b$
and mass low enough to yield a potentially measurable rate for
$\Upsilon$ decay to photon plus $E$-boson, with $E$-boson decay to
$\tau^+\tau^-$ being more likely than decay to the much more difficult
$jj$ final state.  

Such searches have great importance given the fact that $\Upsilon$
decays may be the only way prior to the construction of a linear
collider to obtain confirmation of the existence of $E$-bosons that is
independent of their hoped for observation at the LHC in $h$ decays.
Indeed, generally speaking some light $E$-bosons might appear in $\Upsilon$
decays that do not appear in $h$ decays!

\section{IMPLICATIONS FOR THE LHC \label{sec:lhc}}
The nonstandard cascade Higgs decay scenario has many interesting LHC
implications. Earlier studies have considered whether or not a scalar discovered at the LHC is, in fact, the ``Higgs'' \cite{Burgess:1999ha}. In the present situation, we are interested in cases where new, possibly unexpected decays have arisen, and the consequences for experimental searches. In this review, we attempt to summarize two aspects of
this phenomenology.  For simplicity, we presume the
model predicts one and only one SM-like Higgs boson  and we will refer to
it as the ``Higgs''. The first aspect is the changes in Higgs
phenomenology.  One primary implication for the Higgs is that the
Standard Model decays are subdominant, rendering the LHC Standard Model Higgs searches less effective.  In this regard, it is useful to
consider if the nonstandard Higgs decays will lead to a viable Higgs signal.
However, it is also interesting to discuss potential model-dependent
effects outside of Higgs physics, since such effects combine with the
Higgs as a window into new physics.  The example we will discuss are
changes in the decays of supersymmetric partners of the Standard
Model.  This suggests that a nonstandard Higgs decay could be
accompanied by nonstandard superpartner decays, giving correlated
evidence for this model.

In some regions of parameter space, the intermediate particles facilitating the Higgs cascade decay can have highly displaced vertices.  In this case, the LHCb detector, with its superior ability to trigger on displaced vertices, can have a greater reach for both Higgs and superpartner decays.  Thus, it is even possible that LHCb will be the first LHC experiment to discover this new physics.     


\subsection{Higgs \label{sec:higgssearch}}
The nonstandard Higgs scenario unambiguously predicts changes in the
phenomenology of the SM-like Higgs boson of the model.  At the LHC, the immediate impact is the weakening
of Higgs searches that depend on the Standard Model decay modes.  A
light Higgs near the LEP2 bound is already a difficult region for the
LHC to probe.  Instead of depending on the dominant decay of the Higgs
to $b$ quarks, the experiments have focused on di-photon Higgs decay
and the di-tau decay for a vector boson fusion (VBF) produced Higgs.
Both of these searches are statistics limited, so any suppression of
the Standard Model decay branching ratio increases the required
integrated luminosity for discovery.  Naively, the increase in
required luminosity is a factor of $1/\br(h\to SM)^2$ more.  This
factor is naive as the searches can become more efficient as the
experiment runs, but also the backgrounds can change as the experiment
moves to design luminosity.  At any rate, for a nonstandard Higgs
that is 100 GeV in mass, the Standard Model LEP search determines that
the Standard Model branching ratio is at most 25\%.  Thus, the
required integrated luminosity goes up by a factor of at least 16.
Extrapolating the numbers shown in the CMS TDR \cite{cmstdr}, an
integrated luminosity $\gsim 16*25 \fbi = 400 \fbi$ is needed for
discovery, a significant amount of luminosity.  In the NMSSM context,
a more typical $\br(h\to SM)$ is $\sim 0.1$, implying a need to
increase the luminosity by a factor of roughly 100, as would only be
achievable at the SLHC (assuming no change in the signal to background
ratio --- in fact, $S/B$ will decrease because of the huge number of
multiple interactions). Therefore, in order to maintain the reach for
Higgs discovery it is important to consider whether the nonstandard
Higgs decays can provide a viable Higgs signal.

We focus on the case where there is a SM-like Higgs of the extended
model. Since it couples to Standard Model particles with normal
strength, its production cross sections are unmodified.  So, to
determine the Higgs signal topology we need only specify the
nonstandard decay.  Before proceeding, it is useful to consider the advantages LHC has
over LEP2 to motivate the LHC search strategies.  As a hadron
collider, LHC has higher integrated luminosity and production cross
sections than LEP2 and thus will produce far more of the SM-like
Higgses.  This allows the LHC to look for the rarer clean decay modes
that are not swamped by QCD backgrounds.  Indeed, this is the story
for the Standard Model Higgs.  LEP2 had to look for the dominant
decays of $b \bar b$ and $\tau \bar \tau$, while the LHC instead
searches for $\gamma \gamma$ and for $\tau \bar \tau$ in vector boson
fusion.

\subsubsection{HIGGS DECAYS TO SCALARS}
Higgs decays to a pair of scalars or pseudoscalars (we will use the
notation $a$
in our discussion) is the best known
nonstandard Higgs phenomenology, having been searched for at LEP2 in
the CP-violating MSSM and studied extensively in the NMSSM.  In these
SUSY scenarios, an $\a$ with $\ma>2\mb$ decays 
with branching ratios similar to the SM Higgs boson
and thus $\a\to b\anti b$ decays are dominant with $\a \to \tau\anti
\tau$ being subdominant.

The dominant decay mode of $h\to aa\to 4b$ has strong constraints from
LEP2 which require the Higgs mass to be above 110 GeV.  Still, it is
interesting to see if this decay is capable of being seen above the
QCD background at the LHC for such heavy Higgses.  In recent papers
\cite{Cheung:2007un,Carena:2007jk}, such Higgs decays were studied at
Tevatron and LHC.  These analyses focused on Higgses produced in
association with a $W$ boson and looked at the topology of $4j\, l\,
\nu$, requiring three or four $b$-tagged jets.  For a 120 GeV Higgs,
Ref.~\cite{Carena:2007jk} finds that 5$\sigma$ discovery at the LHC
requires about $30 \fbi$, but is highly reliant on $b$-tagging
efficiencies of 50\% at $p_T \sim 15 \gev$.  Such a high efficiency at
this transverse momentum may be difficult to achieve at the LHC; if this tagging efficiency only holds for 
 transverse momenta greater than 30 GeV, the necessary luminosity goes to $80
\fbi$.  As one extrapolates to lighter Higgses, these issues will
become more important, further increasing the luminosity required.
Their work also studied the $2b 2\tau$ mode and found that it's
prospects were not as promising.  For older work on the $4b$ and $2b\,
2\tau$ modes, see
\cite{Ellwanger:2005uu,Ellwanger:2004gz,Ellwanger:2003jt,Moretti:2006hq,Stelzer:2006sp,Cheung:2007sva}.

If the $\a$ mass is below the $b \bar b$ threshold ($\sim 12 \gev$), the dominant Higgs decay is into $4\tau$ and was only weakly constrained at LEP2.  Thus, at the LHC, it is important to analyze the ideal $m_h\sim 100\gev$ scenario in the $h\to 4\tau$ 
final state.  The two most promising production possibilities are 
vector boson fusion and diffractive Higgs production.  In VBF, one looks for
$WW\to h\to 4\tau$, with tagging of the forward jets emitting the $W$'s in
order to isolate the signal. Studies of this mode have begun.  In diffractive Higgs production, one looks for a
  special class of events with protons appearing in specially designed
  detectors and very little additional activity in the final state.
In a recent paper~\cite{Forshaw:2007ra}, it is claimed that by using
  a track-based analysis in which all events with more than 6 tracks
  in the central region are discarded, a viable signal is possible after
  accumulating $300\fbi$ of integrated luminosity. This type of
  track-based approach may also prove key to extracting a viable
  signal in the VBF fusion channel.
There is also an older analysis, looking for $h\to 4\tau$ at the Tevatron \cite{Graham:2006tr}.

It is perhaps important to mention a particularly useful technique
regarding mass reconstruction in the $4\tau$ case.  Because $\ma\ll
m_h$, the two $a$ decays result in two highly boosted $2\tau$ pairs,
and each pair will decay more or less collinearly to the visible
$2\tau$ decay products and some missing momentum. In the collinear
approximation, there are enough constraint equations to solve for {\it
  both} the $h$ and $a$ masses. In the $WW\to h$ fusion case, this
requires that the $h$ have significant transverse momentum as measured
by the recoiling jets.  In the $pp\to pp h$, the forward tagged
protons actually provide an over-constrained system even if the $h$ has
very little transverse momentum or if this transverse momentum cannot
be well measured.  For details of the latter
see~\cite{Forshaw:2007ra}.

Another class of possible decays arise when $\a$ is fermiophobic to
Standard Model fermion decays \cite{Dobrescu:2000jt,Chang:2005ht}.
This possibility does not arise in the NMSSM, but can occur in other
models.  If $\a$ couples to SM-singlet heavy fermions that it cannot
decay into, its leading decay is through loop-induced decays into
gluons or photons.  Thus, the decay modes of the Higgs are into $4g,\,
2g\, 2\gamma,\, 4\gamma$, in decreasing order of dominance.  The
four gluon decay suffers from too large of a QCD background to be
searched for at the LHC.  However, LEP2 constraints on the decays with
photons could allow $\a\to \gamma\gamma$ branching ratios as high as
$10^{-2}$ \cite{Chang:2006bw}; in this case, the subdominant decays might provide a
viable LHC signal.  In \cite{Martin:2007dx}, the $2g\, 2\gamma$ decay
was analyzed, which showed that, with integrated luminosity of order
$300 \fbi$, a branching ratio for $h \to 2g\, 2\gamma$ of a few
percent was needed to discover the Higgs and $\a$.  In
\cite{Chang:2006bw}, the $h\to 4\gamma$ decays were analyzed for $300
\fbi$.  The background was shown to be negligible and a branching
ratio of $10^{-4}$ for $h\to 4\gamma$ was sufficient to discover both
scalars in most of the parameter space.  A crucial issue for these
fermiophobic decays is efficient triggering.  The photons are
relatively soft, with $p_T \sim m_h/4$, so passing the di-photon
trigger is one of the biggest issues for the signal efficiency.  For
the $4\gamma$ decay, this can be relieved by implementing a multiple
photon trigger with a lower threshold than the di-photon trigger.




\subsubsection{HIGGS DECAYS WITH DISPLACED VERTICES \label{sec:displaced}}
In some cases, the Higgs decays have significantly displaced vertices,
which makes LHCb the ideal detector to search for this Higgs.  Such
Higgs decays have been discussed in supersymmetric models with
R-parity violation \cite{Carpenter:2006hs}, in hidden valley models
\cite{Strassler:2006ri}, and in models with light right-handed
neutrinos \cite{Graesser:2007yj,Graesser:2007pc}.  Displaced vertices
are possible since the intermediate particle, which facilitates the
Higgs cascade decay, can have a suppressed coupling mediating its
decay.  Decay lengths between $100 \,\mu$m and $10$ m are the most
interesting since they are resolvable within the detector.  The LEP2
constraints on such a scenario are difficult to ascertain, especially
for highly displaced vertices.  However, at any rate, such nonstandard
Higgs decays could occur and it is important to determine if there are
ways to detect a Higgs decaying in this fashion.

For the case of R-parity violation that is baryon number violating
\cite{Carpenter:2006hs}, the Higgs decays into a pair of neutralinos
which themselves decay into three quarks each.  If the neutralino
decay is highly displaced, the hadrons formed from the three quarks
point back to a vertex.  So there is a potential of having two
highly displaced vertices that are inconsistent with the Standard
Model.  Such vertices are easiest to detect and trigger on at LHCb,
the dedicated detector for B physics at the LHC.  This detector is
forward focused and thus non-hermetic, but it is possible for the
Higgs to be boosted enough that both neutralino decays occur in the
detector.  An analysis at LHCb \cite{Kaplan:2007ap} showed that such
double displaced vertices have negligible background.  Furthermore,
such Higgs decays produce 1000's of such double displaced events in a
year of LHCb running (integrated luminosity $\sim 2 \fbi$), as long as
the neutralino mass is not too light.  This type of analysis should also apply to other Higgs decays with double displaced vertices as they are only distinguished by the objects related to each vertex.  Those with right-handed neutrinos \cite{Graesser:2007yj,Graesser:2007pc} have dominantly two light quark jets and a charged lepton at a vertex while Hidden Valley models \cite{Strassler:2006ri} typically have $b\anti b$ or $\tau \anti \tau$ at each vertex.  Thus, depending on whether or
not ATLAS/CMS can trigger/isolate such events, it is quite possible
that such Higgses will be discovered first at LHCb.

\subsubsection{HIGGS DECAYS WITH MISSING ENERGY \label{sec:missing}}
Recently, Higgs decays that have both missing energy and visible
energy have been discussed
\cite{Chang:2007de,deGouvea:2007uz,Graesser:2007yj,Graesser:2007pc}.
If the Higgs decays into two new neutral particles, one unstable and
one stable, the final state will contain both missing and visible
energy.  There are many potential decay topologies, but the most
promising ones for searches have two charged leptons in the final
state, giving $l^+ l^- \met$.  Depending upon whether or not the Higgs
is produced with associated particles, the signal topologies look very
similar to standard supersymmetry discovery channels with leptons, for
example dilepton and trilepton events with missing energy
\cite{Chang:2007de}.  The resemblance brings up interesting analysis
issues.  If this Higgs appears in a supersymmetric theory, it will be
necessary to produce cuts to isolate the Higgs component from the
superpartner production component (for example, see
\cite{Baer:1992kd}).  However, a Higgs decaying in this manner can
also appear in a non-supersymmetric theory.  These issues suggest that
to discover such a Higgs could require cooperation between Higgs and
supersymmetry experimentalists --- any analysis which yields an
unexpected excess or one that is inconsistent with an expected signal
should be closely scrutinized.

In these scenarios, the Higgs can also decay invisibly into some
combination of neutrinos and other stable particles.  Searches for
such invisible Higgs decays are already planned for the LHC, where it
has been shown that relatively small values (perhaps as small as
$5-10\%$) for $\br(h\to \met)$ can yield a viable signal in the
$WW$-fusion production mode, assuming SM-like $hWW$
coupling~\cite{Eboli:2000ze,DiGirolamo:2001yv}.  In
\cite{Barger:2006sk}, the potential of using this channel alone to
discover the Higgs was studied in several extended supersymmetric
Standard Models.  The $pp\to pph\to pp \met$ forward diffractive
production channel is also expected to yield a viable signal at full
$300\fbi$ integrated luminosity for $\br(h\to\met)$ significantly
below 1~\cite{Belotsky:2004ex}.  Generically speaking, Higgs decays
containing missing energy can appear in several channels.  Hopefully,
some combination of these searches (and potentially combined with the
Standard Model searches) can make possible the discovery of such a Higgs.

\subsection{Supersymmetric Particle Searches \label{sec:susysearch}}

Aside from changes in Higgs phenomenology, there are important
implications of nonstandard Higgs models for other sectors of the
theory.  First, we have seen that in order to avoid fine-tuning, low
masses for the superpartners (in particular for the stop and the gluino) are
required, typically just beyond current Tevatron limits.  
For the required masses, production rates of superpartners
at the LHC will be very large. Supersymmetry will be discovered with
relatively little integrated luminosity, whereas Higgs discovery will
require large integrated luminosity and will therefore take more time.
Thus, the LHC experimental collaborations should be on the watch for a
situation in which they have discovered supersymmetry but have not
seen any of the Higgs signals in the MSSM for the expected amounts of
integrated luminosity. Higgs channels with cascade decays should then
become a high priority.  In addition, it will be important to
determine whether $WW$ scattering is, or is not, perturbative
in nature.  If it is perturbative, then it is necessary for there to
be one or more relatively light (below $300\gev$) Higgs boson with
large $WW$ coupling that must be searched for.

Implications for BSM particle searches are of even greater importance
in models where the Higgs decays into particles that transform under a
new symmetry.  In supersymmetric models, this symmetry is $R$-parity.  
Supersymmetric particles, being odd under
R-parity, have to cascade decay down into the lightest particle
that is $R$-parity odd, aka the lightest supersymmetric particle
(LSP).  When the (SM-like) Higgs decays into supersymmetric particles, these
will typically be the lighter supersymmetric particles.  The branching
ratio for such decays will often determine or at least constrain the
properties of the light supersymmetric particles, which in turn
constrains how the heavier supersymmetric particles cascade down to
the LSP.

The simplest possibility, where the light ($\leq 100\gev$) Higgs 
decays into two LSPs, is not allowed (without $R$-parity violation)
since the decay is invisible and ruled out by the LEP invisible Higgs search.  
The next simplest possibility is for the Higgs to decay into
the LSP and a heavier supersymmetric particle.  Both particles are
neutral and the heavier one is unstable, decaying down into the LSP.
This gives a missing and visible energy component to the Higgs decay,
which was discussed earlier in Section~\ref{sec:missing}.

In supersymmetric theories, this nonstandard decay can be into
neutralinos or sneutrinos.  For neutralinos, requiring the Higgs to
decay into $\wtil\chi_1 \wtil\chi_0$ and imposing the constraints on
charginos and neutralinos specifies the neutralino spectrum.  This
requires an NMSSM-like supersymmetric theory where the LSP
$\wtil\chi_0$ is mostly singlino in gauge eigenstate composition,
while $\wtil\chi_1$ is mostly bino (as verified via a scan using NMHDECAY
\cite{Ellwanger:2004xm}; see also \cite{Barger:2006kt}).  Thus, the
LSP has very weak couplings mostly mediated through the superpotential
interaction $\what S \what H_u \what H_d$.  In terms of the cascade
decays of supersymmetric particles, this has a drastic consequence.
Their cascades get lengthened, as they will dominantly cascade first
to $\wtil\chi_1$ which will then decay down to $\wtil\chi_0$.  Thus,
at first order, the supersymmetric cascades are as in the MSSM,
followed by a final decay down of the binos into a lighter neutralino.
NMSSM-like models with an extra $U(1)$ and several extra Higgs fields
transforming differently under the extra $U(1)$ can lead to a
complex scenario for Higgs decays and supersymmetric
cascades~\cite{Han:2004yd}. If such a model is realized in nature,
it could take decades to sort out this physics. 

For now, we focus on the less complex nonstandard decay cascades.
Even these have important consequences for collider searches
\cite{Strassler:2006qa,Ellwanger:1998vi}.  At the Tevatron and LEP,
searches for squarks and staus will have weaker constraints due to
these decays \cite{UsandDavid}. At the
LHC, most cascades of superpartner decays will include the
characteristic decay of $\wtil\chi_1$ to $\wtil\chi_0$.  Furthermore,
since the $\wtil\chi_1$ and $\wtil \chi_0$ must be light in order to
appear in the Higgs decay, their production cross sections (at least
that for $\wtil\chi_1\wtil\chi_1$) at the LHC will be large.  Most
probably, other superpartner particles will also be relatively light.
In this case, supersymmetry will be discovered with an early amount of
luminosity ($\sim 30 \fbi$), while the Higgs, because of its
nonstandard decays, will not have been discovered.  By analyzing the
supersymmetry events, it may be possible to measure the branching
ratios of the decays of $\wtil\chi_1$ to $\wtil\chi_0$ and some of the
properties of the $\wtil \chi_1$ and $\wtil \chi_0$. This will help
determine whether the Higgs will have a large $\br(h\to
\wtil\chi_1\wtil\chi_0)$ and provide crucial information regarding the
nonstandard decay topology of the SM-like Higgs.  This information
can then be used to design searches to pick out this decay in the
design luminosity run of the LHC.

In the cascade decay scenario where the neutralino decays via
$R$-parity violation with a displaced vertex \cite{Carpenter:2006hs},
LHCb's capabilities make it possible to efficiently search for
such vertices.  In \cite{Kaplan:2007ap}, the production of a squark
decaying into the lightest neutralino was considered.  Compared to the
Higgs signal, it is less likely for two squarks to appear in LHCb.
Still, there is a reasonable region of parameter space that allows the
discovery of the squark at LHCb, via the appearance of one such
displaced vertex within one year of running.  Thus, it is
possible that LHCb will discover both the Higgs and supersymmetry
before ATLAS/CMS.

\section{CONCLUSION \label{sec:conclusion}}

The motivations for a light ($\leq 100\gev$) Higgs boson with SM-like
couplings to SM particles, but with dominant decays to non-SM
particles, are very substantial.  Typically, and explicitly in the
NMSSM, fine-tuning can be minimized and/or eliminated if the Higgs is
this light.  In addition, a Higgs mass below $100\gev$ is most
consistent with precision electroweak data.  Furthermore, because of
the very small $h\to SM$ decay widths for a light $h$, the $h\to SM$
branching ratios are easily greatly suppressed by the presence of
couplings to a pair of particles (with summed mass below $m_h$) from a
nonstandard sector. Dominance of nonstandard decays typically imply
that LEP limits on $m_h$ are reduced to below roughly $90\gev$ so long
as the ultimate final state is not $h\to 4b$ (for which LEP requires
$m_h>110\gev$). And, in many cases the $h\to b\anti b$ branching ratio
is reduced to the $\sim 10\%$ level that would explain the excess in
$Z+b\anti b$ seen at $M_{b\anti b}\sim 98\gev$ at LEP.

The NMSSM scenarios with no fine-tunings, where the lightest CP-even
Higgs, $\hi$, has $\mhi\sim 100\gev$ and is SM-like in its couplings
to SM particles but decays via $\hi\to \ai\ai\to 4\tau$ or (less
likely, $4j$) and/or $\hi\to \wtil\chi_1\wtil\chi_0\to f\anti f+\met$,
deserve particular attention.  In the NMSSM, to have low fine-tuning
the stop and gluino should be just above Tevatron limits and easily
discoverable at the LHC. If $ \hi\to \wtil \chi_1\wtil\chi_0$,
$m_{\wtil\chi_1}+m_{\wtil \chi_0}<\mhi$ implies
$\wtil\chi_1\wtil\chi_1$ detection may also be possible early on. But,
Higgs discovery will be challenging in all these cases. If
supersymmetric particle decays indicate that $R$-parity is violated
baryonically, then one should also be alert to the possibility that
$\hi\to \wtil\chi_0\wtil\chi_0\to 6j$ decays could be dominant. This
kind of mode can be present in the MSSM as well as the NMSSM. If one
is willing to accept a significant but not outrageous $6\%$ fine
tuning, many more Higgs scenarios emerge. For example, in the NMSSM
the $\hi\to \ai\ai \to 4b$ and $2b+2\tau$ channels with $\mhi>110\gev$
(from LEP limits) would provide possible discovery modes. 

To summarize, the allowed nonstandard Higgs decay topologies are of a few limited types.
There are decays into four SM fermions, which are often mediated by a scalar $\phi$, giving $h\to 2\phi \to 4f$.
Heavier fermions $f$ are usually favored, although the strong limits
on $4b$ decays suggest that $\phi$ is lighter than the $b\anti b$
threshold so that the dominant decay is into $4 \tau$'s.  If the
scalar $\phi$ is fermiophobic, loop-induced decays can generate the
decay $h\to 2\phi \to 4V$, where $V$ is a photon or gluon.  When there
are more than one new state into which the Higgs can decay, it opens
up the possibility of one of these states being stable.  Thus, there
can be decays with both missing and visible energy, usually of the
type $h\to (X_2) X_1 \to (f \bar{f} X_1) X_1 = f\bar{f} + \me$.  In
addition, there is the potential of displaced vertices, when the decay
$\phi$ or $X_2$ are long lived.  These vertices could give LHCb an
inside track on finding such Higgs decays.  Finally, adding additional
particles can make the Higgs cascade decay longer and more complex.
In some cases, input from $B$ factories and/or non-Higgs LHC searches
can pin down some properties of these intermediate states appearing in
the Higgs cascade.  Thus, if no SM Higgs has been found at the LHC at
the expected integrated luminosity, it will be advantageous to use all
available information about these new states to design efficient searches for the nonstandard Higgs decays.             

In general, Higgs decays and phenomenology provide an unexpectedly
fertile probe of and window to physics beyond the Standard Model and
potentially beyond the minimal supersymmetric model. Nonstandard
decays are a double-edged sword, on the one hand possibly making Higgs
detection at the LHC much more difficult while on the other hand
providing information regarding a new sector of the theory.  The
additional particles from the beyond the SM or beyond the MSSM sector
could have very weak direct production cross sections at the LHC and
might only be observed via Higgs decays. Apart from any other
consideration, a highly detailed understanding and delineation of all
Higgs decays will be crucial to understanding BSM physics.  We must
hope that the LHC will prove up to the task, but cannot rule out the
possibility that a linear collider will ultimately be necessary --- at
the very least it would greatly refine the LHC observations.

\section*{ACKNOWLEDGMENTS}
SC and NW are supported
by NSF CAREER grant PHY-0449818 and the U.S. Department of Energy under
grant DE-FG02-06ER41417.
RD is supported by the U.S. Department of Energy 
under grant DE-FG02-90ER40542.
JFG is supported by the U.S. Department of Energy under grant No. DE-FG03-91ER40674.

\bibliography{higgsreview}
\bibliographystyle{h-physrev3}

\end{document}